\documentclass[dvips,12pt,aps,nofootinbib]{revtex4-1}
\usepackage{amsmath,amssymb,epsf,epsfig,amsthm,bm,graphicx,subfigure}

\theoremstyle{theorem}
\newtheorem{theorem}{Theorem}[section]

\newtheorem{proposition}[theorem]{Proposition}
\newtheorem{corollary}[theorem]{Corollary}

\theoremstyle{definition}

\begin{document}

\author{Steven Willison}
\affiliation{Centro de Estudios Cient\'{i}ficos (CECS), Arturo Prat 514, Valdivia, Chile}
\title{The BTZ spacetime as an algebraic embedding}

\begin{abstract}
 A simple algebraic global isometric embedding is presented for the nonrotating  BTZ black hole and its counterpart of Euclidean signature.
The image of the embedding, in Minkowski space of two extra dimensions, is the interection of two quadric hypersurfaces.
Furthermore an embedding into $AdS_4$ or $H_4$ is also obtained, showing that the spacetime is of embedding class one with respect to maximally symmetric space of negative curvature.
The rotating solution of Euclidean signature is also shown to admit a quadratic algebraic embedding, but seemingly requires more than two extra dimensions.
\end{abstract}

{\small Preprint no. CECS-PHY-10/12}

\maketitle

\section{Introduction}
An isometric embedding is an invertible diffeomorphism of a manifold ($M$, $g$) into a submanifold of some higher dimensional space - typically flat space $E^n := (\mathbb{R}^n, \delta)$\cite{Han} or Minkowski spacetime $\mathbb{M}^{p,q} :=(\mathbb{R}^{p+q}, \eta)$ \cite{Greene-70}
- such that the pullback metric induced by the embedding agrees with the  intrinsic metric $g$.
The idea of gaining further understanding of solutions of Einstein's equations in this way
has a long history. For example, an embedding of the Schwarzschild solution into 5+1 dimensional Minkowski space was found by C. Fronsdal\cite{Fronsdal:1959zza} in the 1950s. 
Embeddings are important in the definition of quasilocal mass in General Relativity\cite{Wang}, and have shed light on fundamental questions such as the positive mass theorem and the Riemannian version of the Penrose conjecture. 
Also some features of BPS solutions of M theory/supergravity have been
studied using embeddings into Minkowski space with more than one time dimension\cite{Andrianopoli:1999kx}.

In many ways the simplest example of a black hole is the solution of Ba\~{n}ados, Teitelboim and Zanelli (BTZ)\cite{Banados:1992wn,BHTZ}
 in 2+1 dimensional GR with negative cosmological constant. This can be regarded as a quotient space of Anti de Sitter space and therefore the hyperboloid model in $M^{2,2}$ provides us with a local embedding. For some purposes this is sufficient. (e.g. in  Ref. \cite{Deser:1998xb} a concrete relation was established between the temperature and entropy obtained
from Hawking radiation with that of the Unruh effect for the accelerating observer in the embedding Minkowski space.
The embedding of AdS$_3$ was used, this being sufficient for their purpose, since only the geometry of the ($r,t$) plane was relevant. Embeddings of the $r,t$ plane of other black holes were obtained in \cite{Banerjee:2010ma}.)
However for other purposes it is still desirable to have a faithful global embedding\footnote{If $M$ is the whole (maximally extended) specetime we shall say that the embedding is global. This is appropriate for studying the global and causal structure of the spacetime. If there is a singularity, we regard it as a boundary of $M$. For some well behaved kinds of singularities, the embedding may extend to an injection of $\bar{M}$. This is in fact what happens for the BTZ spacetime (replacing $T>0$ with $T\geq 0$ in Proposition \ref{prop} below), whose singularity is of the same character as the tip of a cone.}, which respects the periodicity of the $\phi$ coordinate.

Here we shall present a global isometric embedding for the BTZ spacetime.
We concentrate mainly on the nonrotating spacetime, which takes the simple form of a real algebraic variety in Minkowski space. The image of the embedding is given by a pair of quadratic equations in the Minkowski space coordinates. This allows a very explicit geometrical representation of the black hole, allowing us to obtain all the geometrical information by elementary algebra, and gives some insight into the nature of the past and future singularities inside the event horizon.

Furthermore, we show that the Euclidean nonrotating solution admits a global isometric embedding into hyperbolic space $H_4$. We analise this using the Klein model of $H_4$  and obtain the donut model described in Proposition \ref{Cylinder_Model}.

In the case of hyperbolic space, all kinds of useful representations and coordinate systems can be obtained from the hyperboloid model by projections. Similarly, the embedding of the nonrotating BTZ spacetime can be used as a unifying model for obtaining many other useful representations. Therefore, in what follows, we will give explicit coordinate transformations to some of the more common representations.

\section{The Euclidean spacetime}

The Euclidean BTZ spacetimes\cite{Carlip:1994gc} are a 2-parameter family of smooth hyperbolic manifolds.
They are obtained from the black hole solutions\cite{Banados:1992wn} by analytic continuation of the standard time coordinate and of the angular momentum parameter. As such they play a role in understanding the thermodynamics of black holes and possibly in quantum gravity, such as strings in three dimensions\cite{Maldacena:2000kv}.
Let us first consider the spacetime without angular momentum, with a single parameter $a$ (proportional to the mass).
A standard way to express the metric is the following:
\begin{gather}\label{EBTZ}
 ds^2 = (\rho^2 -a^2)d \tau_E^2 + \frac{d\rho^2}{\rho^2 - a^2} 
+ \rho^2 d\phi^2\, .
\end{gather}
The range of the coordinates is $a\leq\rho<\infty$, with $\phi \sim \phi +2\pi$ and $\tau_E \sim \tau_E + 2\pi/a$. The latter identification is necesary to avoid the occurence of a conical singularity at $\rho = a$. This periodicity in the Euclidean time
also occurs for the Schwartzschild solution and naturaly lends to the Euclidean path integral an interpretation of a statistical partition function.

There are some geometrical features of the spacetime which are not immediately apparent from formula (\ref{EBTZ}). One sees the global isometries generated by $\partial_\phi$ and $\partial_\tau$ but for example, it is not manifest that the spacetime is a hyperbolic manifold, that is, of constant curvature. There are various alternative ways of writing the metric, which bring to light different geometrical features. Here we shall introduce the global embedded model, which, although straightforwardly obtained,  seems to be new or at least not to have received attention in the literature.
Let us first state the result, and then relate the embedding to some other well known models of BTZ and hyperbolic space.

Let us consider (4+1)-dimensional Minkowski space, with metric written in the form
$ds^2 = -dT^2 + dX^2 + dY^2 +dZ^2 +dW^2$.
\begin{proposition}\label{Euclidean_Result}
 The  non-rotating Euclidean BTZ spacetime can be globally isometrically embedded into  the half-space  ($T>0$) of (4+1)-dimensional  Minkowski space $\mathbb{M}^{4,1}$. The image is the intersection of
 the two hyper-surfaces
$X^2 + Y^2  + Z^2+W^2  -T^2= -1$ and $Z^2 +W^2 = \frac{a^2}{1+a^2}T^2$.
\end{proposition}
The first hyper-surface is nothing but the standard embedding of hyperbolic space. Therefore we have another interesting result:
\begin{corollary}
 The  non-rotating Euclidean BTZ spacetime can be globally isometrically embedded into 4-dimensional
 hyperbolic space ${H}_4$.
\end{corollary}

As proof of Proposition \ref{Euclidean_Result} it is sufficient to give the map $f: M \to \mathbb{M}^{4,1}$, $f:(\rho,\phi,\tau_E) \to (T,X^i)$.
\begin{align}
 T & = \frac{\sqrt{1+a^2}}{a}\rho\, , 
\\
 X & = \frac{\sqrt{\rho^2-a^2}}{a}\cos(a\tau_E)\, , \quad 
Y = \frac{\sqrt{\rho^2-a^2}}{a}\sin(a\tau_E)\, ,
\\
Z & = \rho \cos\phi\, , \qquad 
W = \rho \sin \phi\, .
\end{align}
This map is manifestly injective, with $T$, $X$ and $Y$ being monatonic functions of $\rho$ in the entire range
$a\leq\rho<\infty$ and the periodicities of $\phi$, $\tau_E$ being respected. Also, one can check that the image is a submanifold (to check that the tangent space map $f^*$ is injective at the coordinate singularity $\rho =a$ one can pass to the Kruskal coordinates below). The pullback of the Minkowski metric with respect to $f$ 
is (\ref{EBTZ}). Therefore $f$ is the desired global embedding.

Now let us consider some other models of the Euclidean BTZ and give the mappings which take them into the embedded
model.
\paragraph{The Kruskal coordinates}
A model analogous to the Kruskal model of the Schwartzschild spacetime expresses the metric as:
\begin{gather}\label{EuclidianKruskal}
 ds^2 = 4\frac{dx^2 +dy^2 }{(1-x^2 - y^2)^2}
 + a^2\frac{(1+x^2 + y^2)^2}{(1-x^2 - y^2)^2}
 d\phi^2\, .
\end{gather}
The range of coordinates is $x^2 + y^2 <1$, $\phi \sim \phi+2\pi$. 
This makes it clear that the spacetime is a warped product of the Poincar\'{e}
disk (i.e. the hyperbolic plane) with a circle.
The  embedding is:
\begin{align}
 T(x,y) & =\sqrt{1+a^2}\left(\frac{1+x^2 +y^2}{1-x^2-y^2}\right)\, , \\
X(x,y) &= \frac{2x}{1-x^2-y^2}\, ,\qquad
Y(x,y) = \frac{2y}{1-x^2-y^2}\, ,\\
Z(x,t, \phi) & = a \left(\frac{1+x^2 +y^2}{1-x^2-y^2}\right)\cos \phi\,,\qquad 
W(x,t, \phi)  = a \left(\frac{1+x^2 +y^2}{1-x^2-y^2}\right)\sin \phi\, .
\end{align}

\paragraph{Identification of concentric hemispheres in the upper half space model of ${H}_3$.}
The Euclidean BTZ is a quotient space of ${H}_3$. A nice explicit way to see this, as pointed out in Ref. \cite{Carlip:1994gc}, is using the upper half space model.
Introducing spherical polar coordinates $(\tilde{r}, \tilde{\theta},\tilde{\phi})$ on $\mathbb{R}^3$, we may model ${H}_3$ as the upper half space $0<\tilde{\theta}<\pi/2$ with metric:
\begin{gather}\label{Upper_Half_Space}
 ds^2 = \frac{
      d\hat{r}^2   + \hat{r}^2 d\hat{\theta}^2 +\hat{r}^2 \sin^2 \hat{\theta}d\hat{\phi}^2
                        }
                    {
       \hat{r}^2 \cos^2 \hat{\theta}
                         }
\end{gather}
The manifold is obtained by identifying concentric hemispheres 
 $\hat{r} \sim \hat{r}e^{2\pi a}$.

The embedded model is obtained by the map:
\begin{align}
 T & = \frac{\sqrt{1+a^2}}{\cos\hat{\theta}}\, , 
\\
 X & = \tan\hat{\theta}\cos(\hat{\phi})\, , \quad 
Y = \tan\hat{\theta}\sin(\hat{\phi})\, ,
\\
Z & = \frac{a}{\cos\hat{\theta}}  \cos\left (\frac{ \text{ln}\,\hat{r} }  {a}\right)\, , \qquad 
W =  \frac{a}{\cos\hat{\theta}}  \sin\left (\frac{ \text{ln}\,\hat{r} }  {a}\right) \, .
\end{align}

\paragraph{Multivalued map from hyperboloid model of ${H}_3$ into hyperboloid model of ${ H}_4$}

Let us now represent $H_3$ as the hyper-surface $\xi_0^2 -\xi_i \xi_i =-1$, $i =1,2,3$ in $\mathbb{M}^{3,1}$. One obtains:
\begin{align}
 T & = \sqrt{1+a^2}\sqrt{\xi_0^2 -\xi_3^2}\, , 
\\
 X & =\xi_1\, , \quad 
Y = \xi_2\, ,
\\
Z & = a \sqrt{\xi_0^2 -\xi_3^2}  \cos\left (\frac{1}{a} \text{ln}\left(2\sqrt{\frac{\xi_0-\xi_3}{\xi_0+\xi_3}}\right)   \right)\, , 
\\
W & =   a \sqrt{\xi_0^2 -\xi_3^2}  \sin\left (\frac{1}{a} \text{ln}\left(2\sqrt{\frac{\xi_0-\xi_3}{\xi_0+\xi_3}}\right)   \right)\, .
\end{align}

\subsection{Explicit embedding into four dimensional hyperbolic space}

As stated in the Corollary above, we can embed the manifold into $H_4$. It is therefore interesting to forget the Minkowski space and consider an intrinsic parameterisation of hyperbolic space. Since the embedding in Minkowski is expressed by homogeneous quadratic forms, one can naturally pass to Klein´s projective model. The standard inhomogeneous coordinates are obtained from the hyperboloid model by projection from the origin onto the surface $T =1$ i.e. $(u^i, 1)$, with $u^i := X^i/T$. The entire hyperbolic space is represented as the 
Klein ball $K_4:= \{\bm{u}\, | u^i u^i < 1\}$ with metric function:
\begin{gather}
 d(\bm{u},\bm{u^\prime}) = \frac{1 - u_i u^\prime_i}{\sqrt{1-u_iu_i}\sqrt{1-u^\prime_i u^\prime_i}} \, .
\end{gather}
Rather attractively, the BTZ is embedded as:
\begin{gather}
 (u_3)^2 + (u_4)^2 = \frac{a^2}{1+a^2}\, .
\end{gather}
A nice feature of the Klein ball is that the geodesics are represented as straight lines. The surfaces with $z,w$ constant are totally geodesic and are isometric to the hyperbolic plane. As mentioned above, the BTZ is a warped 
product of the hyperbolic plane with a circle. We now see that this warped product  looks like a cartesian product in the natural cartesian coordinates of Klein's model.

We may regard the Euclidean black hole as a quotient space of the 3 dimensional Klein model with inhomogeneous coordinates $(v_1,v_2,v_3)$. From the map from section c above we obtain a map $K_3 \to K_4$:
\begin{align}
 \sqrt{1+a^2} \, u_1 & =\frac{v_1}{\sqrt{1-(v_3)^2}}\, , \quad 
  \sqrt{1+a^2} \, u_2  =\frac{v_2}{\sqrt{1-(v_3)^2}}\, ,
\\
\sqrt{1+a^2}\, u_3 & = a  \cos\left (\frac{1}{a} \text{ln}\left(2\sqrt{\frac{1-v_3}{1+v_3}}\right)   \right)\, , 
\\
\sqrt{1+a^2}\, u_4 & =   a \sin\left (\frac{1}{a} \text{ln}\left(2\sqrt{\frac{1-v_3}{1+v_3}}\right)   \right)\, .
\end{align}
By considering the distance function for $K_3$ under this map, and introducing new coordinates $x:=\sqrt{1+a^2} \, u_1$, $y:= \sqrt{1+a^2} \, u_2$ and $\phi$ for the angle in the $u_3$, $u_4$ plane, we obtain the following model:

\begin{proposition}[Donut Model of the Euclidean nonrotating BTZ]\label{Cylinder_Model}
Let $C := D_2 \times S_1 = \{\bm{x}:=(x,y,\phi) \, | x^2 +y^2 < 1, \phi\sim\phi+2\pi\}$ with distance function:
\begin{gather}
 \cosh  d(\bm{x},\bm{x}^\prime) = \frac{ \cosh\left(a(\phi -\phi^\prime + 2 n \pi)\right) -xx^\prime -yy^\prime}{\sqrt{1 -x^2 -y^2}\sqrt{1 -(x^\prime)^2 -(y^\prime})^2}
\end{gather}
where $n$ is an integer chosen such that $\phi -\phi^\prime + 2 n \pi$ is in the range $(-\pi,\pi]$. Then $C$ is isometric to the nonrotating Euclidean BTZ black hole.
\end{proposition}
It is quite remarkable that the dependence on $\phi$ and $\phi^\prime$ factorises out of the denominator. Also,  we may easily obtain the geodesics, since they are the images of straight lines in $K_3$. They are of two types:
\\
 i)  curves winding around the circle, whose restriction to the $x$, $y$ plane trace out a hyperbola $x= c_1 \cosh a\phi +c_2 \sinh a\phi$,  $y= c_3 \cosh a\phi +c_4 \sinh a\phi$;
\\
 ii) straight lines in the $x,y$ plane with $\phi$ constant.
\\
Any two points are connected by an infinite number of geodesics corresponding to the infinite number of images $(x,y, \phi +2m\pi)$ of each point in the covering space.

\section{The Lorentzian nonrotating BTZ solution}

To find the embedding of the black hole solution one essentialy follows the same steps as above, replacing $Y \to iS$.
\begin{proposition}\label{prop}
 The nonrotating BTZ black hole spacetime can be globally isometrically embedded into the halfspace $T>0$ of Minkowski space $\mathbb{M}^{2,3}$ as the intersection of the two hyper-surfaces
$- S^2 + X^2 +Z^2+W^2 -T^2= -1$ and $ Z^2  +W^2= \frac{a^2}{1+a^2}T^2$ .
The past and future singularities are located at the intersection of these surfaces with the hyperplane $T=0$.
\end{proposition}

\textbf{The standard coordinates:}
The standard way of presenting the metric is:
\begin{gather}\label{BTZ}
 ds^2 = -(\rho^2 -a^2)d \tau^2 + \frac{d\rho^2}{\rho^2 - a^2} 
+ \rho^2 d\phi^2\, .
\end{gather}
Two copies of the exterior region $\rho > a$ are mapped into the embedded manifold by:
\begin{align}
 T & = \frac{\sqrt{1+a^2}}{a}\rho\, , 
\\
 X & = \pm\frac{\sqrt{ \rho^2 - a^2}}{a}\cosh(a\tau)\, , \quad 
S = \frac{\sqrt{\rho^2-a^2}}{a}\sinh(a\tau)\, ,
\\
Z & = \rho \cos\phi\, , \qquad 
W = \rho \sin \phi\, .
\end{align}
Two copies of the interior region $\rho<a$ are mapped by:
\begin{align}
 T & = \frac{\sqrt{1+a^2}}{a}\rho\, , 
\\
 X & = \frac{\sqrt{a^2- \rho^2}}{a}\sinh(a\tau)\, , \quad 
S = \pm\frac{\sqrt{a^2- \rho^2}}{a}\cosh(a\tau)\, ,
\\
Z & = \rho \cos\phi\, , \qquad 
W = \rho \sin \phi\, .
\end{align}

\textbf{The Kruskal type coordinates}:
The metric in Kruskal-type coordinates is:\cite{BHTZ}
\begin{gather}\label{LorentzianKruskal}
 ds^2 = 4\frac{-dt^2 +dx^2 }{(1+t^2-x^2 )^2}
 + a^2\frac{(1-t^2 +x^2)^2}{(1+t^2 -x^2)^2}
 d\phi^2\, .
\end{gather}
The domain of the coordinates is $-1 <-t^2 +x^2< 1$, $\phi\sim \phi +2\pi$. The singularities
are at $t^2 -x^2 =1$ and conformal infinity is at $x^2 -t^2 =1$. The event horizons are at $x = \pm t$ with bifurcation surface 
at $x=t=0$.
This covers the maximally extended space-time. 
So, to make more explicit the fact that our embedding is global, we give here the map:
\begin{align}
 T(x,t) & =\sqrt{1+a^2}\left(\frac{1-t^2+x^2 }{1+t^2-x^2}\right)\, , \\
X(x,t) &= \frac{2x}{1+t^2-x^2}\, ,\qquad
S(x,y) = \frac{2t}{1+t^2-x^2}\, ,\\
Z(x,t, \phi) & = a \left(\frac{1-t^2+x^2}{1+t^2-x^2}\right)\cos \phi\,,\qquad 
W(x,t, \phi)  = a \left(\frac{1-t^2+x^2}{1+t^2-x^2}\right)\sin \phi\, .
\end{align}
We may now take the $(T,X,S)$ subspace of our embedded model as a kind of 3-dimensional Kruskal diagram. As in the standard diagram, each point is actually a circle (with radius $\rho = aT/\sqrt{1+a^2}$). The diagram consists of that half of the hyperboloid
$S^2 + T^2/(1+a^2)= X^2 +1$ for which $T>0$. The two singulatities are the curves $S^2 = X^2+1$, $T=0$. The horizons are the two straight straight lines $S = \pm X$, $T=\sqrt{1+a^2}$ passing through the bifurcation point $X,S=0$ (the fact that each point on a hyperboloid has two straight lines passing through it was apparently discovered by Sir Christopher Wren).

\textbf{Alternative Global Coordinates:}
In Refs. \cite{Bieliavsky:2002ki} the global structure was studied in terms of Lie theory and foliations of symmetric spaces. They obtained a global expression for the metric of the rotating and nonrotating black hole. The latter, which with a slight change of notation we express as
\begin{gather}
  ds^2 = d\xi^2 + \cosh^2 \xi \left( -d \varphi^2 + a^2 \sin^2 \varphi d\phi^2\right)\, ,
\end{gather}(with $-\infty < \xi < \infty$, $0<\varphi <\pi$ and $\phi \sim \phi +2\pi$)
 is very naturally related to the embedded model as follows:
\begin{align}
 X &= \sinh \xi\, ,\qquad
S = \cosh\xi \cos \varphi\, ,\qquad
T  = \sqrt{1+a^2}\cosh \xi \sin \varphi  \, , \\
Z & = a \cosh\xi \sin\varphi \cos\phi\,,\qquad 
W  = a \cosh\xi \sin\varphi \sin \phi\, .
\end{align}

The embedding offers an interesting new perspective. We see that the continuation of spacetime ``beyond the singularity", by including another copy embedded into the region $T<0$, is as natural as the extension of a cone to a double cone. The continued spacetime would then be two copies of the eternal black hole with past  singularity of one copy identified with the future singularity of the other and vice versa. Hence closed timelike curves, winding around the $S,T$ plane in the embedding space, would pass across the singularities.

Some more comments are in order.
The submanifold $- S^2 + X^2 +Z^2+W^2 -T^2= -1$ has the topology of $S_1\times \mathbb{R}_3$. It is what mathematicians would call Anti de Sitter space. We shall refer to it as $\text{AdS}_4$. It contains closed timelike curves (CTCs) and therefore it is not a physically meaningful spacetime.  Physicists normally use the label to describe the spacetime of constant negative curvature with topology of $\mathbb{R}^4$, which we shall refer to as $\widetilde{\text{AdS}}_4$, obtained by uncompactifying the angular coordinate in the $S$, $T$ plane. An embedding into $\widetilde{\text{AdS}}_4$ is straightforwardly obtained from this.
Since the embedding is in a region in which angular coordinate in the $S$, $T$ plane only runs from $0$ to $\pi$, the embedding is consistent with decompactifying.
Therefore:
\begin{corollary}
 The nonrotating BTZ black hole admits global isometric embeddings into $\widetilde{\text{AdS}}_4$ and $\text{AdS}_4$.
\end{corollary}

Intuitively, the BTZ spacetime is the negative curvature analogue of the conical spacetime of Deser, Jackiw and 't Hooft\cite{Deser:1983tn}, describing a point particle in 2+1 dimensions without cosmological constant. The embedding offers a new way of viewing this.
Let us denote by $\mathcal{C}_2$ a subluminal cone in $\mathbb{M}^{2,1}$ . Then the BTZ spacetime
is the intersection  $(\mathcal{C}_2\times \mathbb{M}^{1,1})\cap\text{AdS}_4$ where these spaces are understood to be embedded into 
 $\mathbb{M}^{3,2}$ in the natural way, as in proposition  
\ref{prop}.
The point particle metric has geometry $C_2\times $ time, where $C_2$ is a Euclidean cone. This can be embedded into  $ \mathbb{M}^{3,1}$. By introducing an extra space dimension in a trivial way we may regard the point particle as $(C_2\times   \mathbb{M}^{1,1}) \cap \mathbb{M}^{3,1}$, embedded into $ \mathbb{M}^{4,1}$ .
In summary:
\\\\
BTZ: $((\mathcal{C}_2\times \mathbb{M}^{1,1})\cap\text{AdS}_4) \subset\mathbb{M}^{3,2}$
\\\\
Point particle:  $((C_2\times  \mathbb{M}^{1,1}) \cap \mathbb{M}^{3,1})\subset \mathbb{M}^{4,1}$ 
\\\\

\section{The Euclidean rotating solution}

Let us now briefly discuss the rotating Euclidean BTZ spacetime. This can be obtained as a quotient 
of the upper half space model of $H_3$ by identifying $(\hat{r},\hat{\phi})\sim (\hat{r}e^{2\pi a},\hat{\phi}+2\pi b)$ in (\ref{Upper_Half_Space}).
Introducing coordinates
$\phi := (1/a )\log \hat{r}$, $\tau_E:= -(b/a^2) \log \hat{r}+ \hat{\phi}/a$, $\rho := a/\cos\hat{\theta}$, then the metric takes the form
\begin{gather}
   ds^2 = (\rho^2 -a^2)(d \tau_E +\frac{b}{a} \phi)^2 + \frac{d\rho^2}{\rho^2 - a^2} 
+ \rho^2 d\phi^2\, ,
\end{gather}
with $\rho \geq a$ and the identifications being simply $\phi\sim \phi+2\pi$, $\tau_E \sim \tau_E +2\pi/a$. 
Note that $a$ and $b$ correspond to the outer and inner horizon radii respectively of the Lorentzian spacetime.
Alternatively, introducing $\chi: = \cosh^{-1}(\rho/a)$, $\theta := a\tau_E$ ( $0\leq \chi <\infty$),
\begin{gather}
 ds^2 = 
 \sinh^2 \chi d\theta^2 + 2b\sinh^2 \chi \, d\phi d\theta
 + \left( b^2 \sinh^2 \chi + a^2 \cosh^2 \chi  \right) d\phi^2 + d\chi^2\, .
\end{gather}

The embedding we present here is only partially  successful  in regard to simplicity, since it involves many extra spacelike and timelike dimensions. But it does preserve the property of being defined only in terms of quadratic equations.
Perhaps a more efficient embedding can be found. This is left as an open problem.
The inspiration for what follows comes from the embedding of a twisted torus in Euclidean space by winding around a higher dimensional untwisted torus. Hence we start by rewriting the metric in the form:
\begin{gather*}
ds^2= |b|\sinh^2 \chi (d\theta \pm d\phi)^2 + (1-|b|) \sinh^2 \chi d\theta^2 + \left(
 a^2 \cosh^2 \chi  -|b|(1-|b|) \sinh^2 \chi\right) d\phi^2 + d\chi^2
\end{gather*}
where $\pm$ is the sign of $b$.
A global isometric embedding  into $M^{3,7}$, with
$ds^2 = -dT^2 -dS_1^2 -dS_2^2 + dX_a dX_a$,
 is parameterised by:
\begin{align}
T & = \sqrt{1+a^2} \cosh^2 \chi\, ,
\\
S_1  & = \sqrt{|b|(1-|b|)} \sinh \chi \cos \phi\, ,
\\
S_2 & =  \sqrt{|b|(1-|b|)} \sinh \chi \sin \phi\,
\\
X_1 & = \sqrt{|b|} \sinh \chi \cos (\theta \pm\phi)\, ,
\\ 
X_2 & = \sqrt{|b|} \sinh \chi \sin (\theta \pm\phi)\, ,
\\
X_3 & = \sqrt{(1-|b|)} \sinh \chi \cos \theta\, ,
\\
X_4 & = \sqrt{(1-|b|)} \sinh \chi \sin \theta\, ,
\\
X_5 & = a \cosh\chi \cos \phi\, ,
\\
X_6 & = a \cosh\chi \sin \phi\, ,
\\
X_7 & =  \sqrt{|b|(1-|b|)} \sinh\chi\, .
\end{align}
(If $|b|>1$ the signature of the embedding space should be changed accordingly i.e. $X_7 \to i X_7$). The global nature of the embedding is guaranteed by the fact that $\sinh \chi$ and $\cosh \chi$ are monotonic over the relevant range of $\chi$.
Using the double angle formula for $\sin(\theta\pm \phi)$ one obtains a set of quadratic equations. Therefore:

\begin{proposition}
The rotating Euclidean BTZ spacetime admits a global isometric embedding  into the region $T,X_7 \geq 0$ of $M^{3,7}$ as an intersection of the quadric hypersurfaces:
\begin{align}
S_1^2 +S_2^2 &= (1-|b|)(X_1^2 +X_2^2) = |b|(X_3^2+X_4^2) = X_7^2 = |b|(1-|b|) \left(\frac{T^2}{1+a^2} -1\right)\, ,
\nonumber\\
X_5^2 +X_6^2 &= \frac{a^2}{1+a^2}T^2\, ,
\\\nonumber
X_4 X_5 \pm X_3 X_6 &= \frac{a}{\sqrt{1+a^2}} \sqrt{\frac{1-|b|}{|b|} } X_2 T\, .
\end{align}
\end{proposition}

It  is straightforward to check that this is a submanifold of the hypersurface $X^i X^i -S^aS^a -T^2 = -1$. Therefore we get an embedding into AdS$_{2,7}$, that is, a generalisation of AdS with two times.

\section{The Lorentzian Rotating Solution}

A standard way of expressing the  general BTZ metric in 2+1 dimensions is:
\begin{gather}
 ds^2 = -dt^2 \frac{(r^2 -r_+^2)(r^2 - r_-^2)}{r^2} + \frac{r^2 dr^2}{(r^2 -r_+^2)(r^2 - r_-^2)}
 + r^2\left(d\psi^2 -\frac{r_+r_-}{r^2} dt\right)^2\, .
\end{gather}
where $r_+$ and $r_-$ are the outer and inner horizons respectively. This describes a rotating black hole with mass and angular momentum given by
$M = r_+^2 +  r_-^2$ and $J = 2r_+r_-$.
By the change of coordinates $\tau = t \frac{(r_+^2 -r_-^2)}{r_+^2}$, $\rho = r_+ \sqrt{\frac{r^2 -r_-^2}{r_+^2 -r_-^2}}$, $\phi = \psi + \frac{r_+}{r_-}t$, we put the metric in the form:
\begin{gather}
 ds^2 = -(\rho^2 -r_+^2)(d\tau + \frac{r_-}{r_+}d\phi)^2 +\frac{d\rho^2}{\rho^2 - r_+^2} + \rho^2 d\phi^2\, . 
\end{gather}
Note that the above coordinate transformation is not defined for the extremal case $r_+ = r_-$.
Following the above steps, we first write
\begin{gather*}
ds^2= |b|\sinh^2 \chi (d\hat\tau \mp d\phi)^2 -(1+|b|) \sinh^2 \chi d\hat\tau^2 + \left(
 a^2 \cosh^2 \chi  -|b|(1+|b|) \sinh^2 \chi\right) d\phi^2 + d\chi^2
\end{gather*}
where $\hat\tau := a\tau$.
By analogy with the Euclidean metric we may try the following:
\begin{align}
T & = \sqrt{1+a^2} \cosh \chi\, ,
\\
X_1 & = \sqrt{|b|} \sinh \chi \cos (\hat\tau \mp\phi)\, ,
\\ 
X_2 & = \sqrt{|b|} \sinh \chi \sin (\hat\tau \mp\phi)\, ,
\\
X_3 & = a \cosh\chi \cos \phi\, ,
\\
X_4 & = a \cosh\chi \sin \phi\, ,
\\
X_5 & = \sqrt{(1+|b|)} \sinh \chi \cosh \hat\tau\, ,
\\
S_1 & = \sqrt{(1+|b|)} \sinh \chi \sinh \hat\tau\, ,
\\
S_2  & = \sqrt{|b|(1+|b|)} \sinh \chi \cos \phi\, ,
\\
S_3 & =  \sqrt{|b|(1+|b|)} \sinh \chi \sin \phi\,
\\
S_4 & =  \sqrt{|b|(1-|b|)} \sinh\chi\, .
\end{align}
However, this does not lead to quadrics. Rather, the double angle formula for $\hat{\tau} \mp \phi$ will lead to a curious expression of the form:
\begin{gather}
 (1+|b|) \left(\frac{X_1 + i X_2}{S_2 +iS_3}\right)^2  = \left( \frac{X_5 +S_1}{X_5 -S_1}\right)^{\mp i}\, .
\end{gather}
It is unclear whether this will give any geometrical insight. Also it is unclear at the present how to proceed for the extremal spacetime.

\section{Negative mass and  Zero Mass spacetime}
We focus on the nonrotating spacetime with Euclidean signature.
The solution of negative mass is described by the line element:
\begin{gather}
 ds^2 = (\rho^2 +  \alpha^2)d \tau_E^2 + \frac{d\rho^2}{\rho^2 +\alpha^2} 
+ \rho^2 d\phi^2\, ,
\end{gather}
with $\alpha \neq 0$.

\begin{proposition}
 The Euclidean BTZ spacetime with negative mass can be globally isometrically embedded into $H_4$. It can be globally isometrically embedded into the region  ($T>0$, $X>0$) of $\mathbb{M}^{4,1}$ as the intersection of
 the two hyper-surfaces
$X^2 + Y^2  + Z^2+W^2 -T^2= -1$ and $Z^2+W^2= \frac{\alpha^2}{1-\alpha^2}X^2$ .
\end{proposition}
The map is:
\begin{align*}
 T &= \frac{\sqrt{\rho^2 +\alpha^2}}{\alpha}\cosh \alpha \tau_E\, ,
\\
 X &= \frac{\sqrt{1-\alpha^2}}{\alpha} \rho 
\\ Y &= \frac{\sqrt{\rho^2 + \alpha^2}}{\alpha} \sinh \alpha \tau_E
\\
Z &= \rho \cos \phi\, ,
\\
W &= \rho \sin\phi\, .
\end{align*}
We have an embedding into a submanifold of $H_4$ which is given by $z^2 +w^2 =  \frac{\alpha^2}{1-\alpha^2}x^2$ in both the Klein Ball and the Conformal ball models. Note that $\tau_E$ is not periodic in this case. There is a conical singularity at $X=0$.

The zero mass solution $\alpha \to 0$ can be expressed, introducing $u := 1/\rho$ as:
\begin{gather}
 ds^2 = \frac{du^2 +  d \tau_E^2 +  d\phi^2}{u^2}\, .
\end{gather}
If $\tau_E$ is not regarded as periodic, this embeds straightforwardly into the upper half space model of $H_4$.
If $\tau_E$ is periodic we may embed into $H_5$.
Let us consider the first case. Note that the period of $\phi$ is arbitrary so we may write $\phi = c \hat{\phi}$ where the period of $\hat{\phi}$ is $2\pi$. 
Passing to the hyperboloid model, we deduce:
\begin{proposition}
 The Euclidean BTZ spacetime with zero mass and non-periodic in Euclidean time can be globally isometrically embedded into $H_4$. It can be globally isometrically embedded into the region  ($T>0$, $X>0$) of  $\mathbb{M}^{4,1}$ as the intersection of
 the two hyper-surfaces
$X^2 + Y^2  + Z^2+W^2 -T^2= -1$ and $Z^2+W^2= c^2(T+X)^2$ .
\end{proposition}
The image of the embedding in the Klein ball $K_4$ is $z^2 +w^2 = c^2(1+x)^2$, which makes clear that there is a ``cusp at infinity" at $x=-1$.

\section{Concluding Remarks}

When looking for an embedding of the Schwarzschild solution, following a standard procedure, one encounters elliptic integrals. We have seen that for the 2+1 dimensional spherically symmetric black hole, things are dramatically simpler.
We have shown that the nonrotating BTZ black hole is globally isometrically embedded into $\mathbb{M}^{3,2}$ as an algebraic submanifold. I would suspect that this is the minimal number of dimensions, but I know of no proof.

We have shown that the Euclidean BTZ spacetime \emph{is} of embedding class one with respect to a hyperbolic space with the same scalar curvature  (and likewise for the Lorentzian black hole w.r.t. Anti de Sitter space).
Focusing on Euclidean signature and including positive, negative and zero mass solutions, we obtain a satisfying unified picture:
In each case, we have the intersection in $\mathbb{M}^{4,1}$ of the hyperboloid model of $H_4$ with a cone. The axis of the cone is timelike, spacelike and null respectively.
 For future study, there naturally arises the question of rigidity: Is the embedding unique up to a global isometry of $H_4$?

Embeddings have been given for the rotating solution and at least in the Euclidean case an algebraic (quadratic) embedding exists.
However, these are not altogether satisfactory and a more economical embedding would be desirable.

\acknowledgements

I would like to thank J. Zanelli for many illuminating conversations on the subject. I also thank David Tempo for a helpful discussion about some similar embeddings of other spacetimes in $2+1$ dimensions, R. Baeza for kindly explaining to me some of the mathematical properties of algebraic varieties and Ph. Spindel for bringing to my attention the work of Refs. \cite{Bieliavsky:2002ki} on global structure of the black hole solution. This work has been partially funded by the Conicyt grant ACT-91:
``Southern Theoretical Physics Laboratory" (STPLab). The Centro de Estudios
Científicos (CECS) is funded by the Chilean Government through the Centers
of Excellence Base Financing Program of Conicyt.

 \end{document}